\documentclass[aps,pra,twocolumn,showpacs,superscriptaddress]{revtex4-1}
\usepackage{mathrsfs}

\usepackage{amsfonts,mathrsfs,amsmath,amsthm,amssymb}
\usepackage{graphicx}
\usepackage{dcolumn}
\usepackage{bm}
\usepackage{times}
\usepackage[colorlinks=true,breaklinks=true,linkcolor=blue,citecolor=blue,urlcolor=blue]{hyperref}

\makeatletter
    
    \newcommand{\Rmnum}[1]{\expandafter\@slowromancap\romannumeral #1@}
\makeatother

\begin{document}

\author{Long Huang}
\affiliation{College of Physical Science and Technology, Sichuan University, Chengdu 610064, China}
\author{Bo You}
\affiliation{College of Physical Science and Technology, Sichuan University, Chengdu 610064, China}
\author{Xiaohua Wu}
\email{wxhscu@scu.edu.cn}
\affiliation{College of Physical Science and Technology, Sichuan University, Chengdu 610064, China}
\author{Tao Zhou}
\email{taozhou@swjtu.edu.cn}
\affiliation{School of Physical Science and Technology, Southwest Jiaotong University, Chengdu 610031, China}

\date{\today}

\title{Generality  of the concatenated five-qubit code}
\begin{abstract}
In this work, a quantum error correction (QEC) procedure with the concatenated five-qubit code is used to construct a near-perfect effective qubit channel (with a error below $10^{-5}$) from arbitrary noise channels. The exact performance of the QEC is characterized by a Choi matrix, which can be obtained via a simple and explicit protocol. In a  noise model with five free parameters,  our numerical results indicate that the concatenated five-qubit code is general: To construct a near-perfect effective channel from the noise channels, the necessary size of the concatenated five-qubit code depends only on the entanglement fidelity of the initial noise channels.

\end{abstract}
 \maketitle

 \section{Introduction}
In quantum computation and communication, quantum error correction (QEC) is necessary for preserving coherent states from noise and other unexpected interactions. Based on classic schemes using redundancy, Shor~\cite{1} has championed a strategy where a bit of quantum information is stored in an entanglement state of nine qubits. This scheme permits one to correct any error incurred by any of the nine qubits. For the same purpose, Steane~\cite{2} has proposed a protocol that uses seven quits. The five-qubit code was discovered by Bennett, DiVincenzo, Smolin and Wootters~\cite{3}, and independently by Laflamme, Miquel, Paz and Zurek~\cite{4}. The QEC conditions were proved independently by Bennett and co-authors~\cite{3}, and by Knill and Laflamme ~\cite{5}. The protocols above with different quantum error correction codes (QECCs) can be viewed as active error correction. There are passive error avoiding techniques such as the decoherence-free subspaces~{\cite{6,7,8} and noiseless subsystem~\cite{9,10,11}. Recently, it has been proved that all the active and passive QEC methods can be unified~\cite{12,13,14}.

With the known codes constructed in Refs.~\cite {2,3,4}, the standard QEC procedure is designed according to the principle of perfect correction for arbitrary single-qubit errors, where one postulates that single-qubit errors are the dominant terms in the noise process~\cite{15}. Recently, an optimization-based approach to QEC was explored. In each case, rather than correcting for arbitrary single-qubit errors, the error recovery scheme was adapted to model for the noise, with the goal to maximize the fidelity of the operation~\cite{16,17,18,19}. Robust channel-adapted QEC protocols, where the uncertainty of the noise channel is considered, have also been developed~\cite{20,21,22}.

For some tasks such as storing or transferring a qubit of information, a near-idealized channel is usually required. If the fidelity obtained from error correction is not high enough, the further increase in levels of concatenation is necessary. In the previous works~\cite{Poulin2,Rahn,Kersting}, the application of QEC with the concatenated code was discussed for the Pauli channel which includes the depolarizing channel as the most important example. In general, a QEC  protocol contains three steps: encoding, error evolution, and decoding. When the concatenated code is used for encoding, there are two known methods for decoding: the widespread blockwise hard decoding technique and the optimal decoding using a message-passing algorithm~\cite{Poulin2}. As shown by Poulin, the Monte Carlo results using  the five-qubit and Steane's code on depolarizing channel reveal significant advantages of message-passing algorithms. For the depolarizing channel, the concatenated five-qubit code is also more efficient than the concatenated seven-qubit code.

In the present work, we shall focus on the following questions:  For an arbitrary noise model, instead of finding the optimal QEC protocol adapted to it, is it possible for us to construct a near-perfect channel with a error below $10^{-5}$ by performing a QEC procedure with the concatenated five-qubit code? In order to answer this question, two important methods developed in the previous works are applied here. At first, following the idea in Ref.~\cite{3}, we shall show that the blockwise decoding can be carried out in a way without performing the error syndrome, and the realization needs some additional quantum resources, required in the message-passing algorithms. The other method comes from the recent works where several  general schemes have been developed for describing the exact performance of the QEC procedure based on a $N^l$-qubit concatenated code~\cite{Rahn,Kersting}. The main ideas in these schemes can be summarized in the following: First, the exact performance of the QEC with a $N^l$-qubit code is denoted by an effective Choi matrix; Secondly, the Choi matrix in the $l$-th level of concatenation is obtained by simulating the standard quantum process tomography (SQPT)~\cite{15,23,24,25,26,27}. Based on these results, the Choi matrix in the $(l+1)$-th level can be obtained in a similar way. The advantage of introducing the effective Choi matrix is clear: Instead of directly working in the $2^{N^{l}}$-dimensional Hilbert space, one could always simulate the error correction in each level of concatenation in a $2^N$-dimensional system.

For a general noise model, with five free parameters, our numerical simulation indicates that the concatenated five-qubit code is general: The QEC protocol with the concatenated five-qubit code, which is able to construct a near-perfect effective channel from the noise depolarizing channels, is also sufficient to complete the same task for other types of noise channels with the same channel fidelity.

The content of present work is organized as follows. In Sec.~\ref{sec2}, we construct an error correction protocol with the five-qubit code~\cite{3}, and a chosen unitary transformation is shown to be sufficient for correcting the errors of the principle system. In Sec.~\ref{sec3}, an explicit scheme is designed to obtain the effective Choi matrix. For three types of channels, the depolarizing channel, the amplitude damping channel and the bit-flip channel, the effective Choi matrices obtained by performing QEC with the concatenated five-qubit code is shown in Sec.~\ref{sec4}. In Sec.~\ref{section5}, it is shown that the concatenated seven-qubit is not general. In Sec.~\ref{sec6}, a noise model containing five free parameters is constructed, and we argue that the concatenated five-qubit code is general. Finally, we end our work with a short discussion.

\section{unitary realization of quantum error correction}
\label{sec2}
The $N$-qubit code concatenated with itself $L$ times yields a $N^{L}$-qubit code, providing a better error resistance with increasing $L$. The direct simulation of the quantum dynamics, coding and encoding procedure require massive computation resources. By following the idea in Refs.~\cite{Rahn,Kersting}, the concept that the exact performance of QEC can be described by an effective channel, will make the calculation simplified.

Let us consider the QEC protocol with code in Ref.~\cite{3},
\begin{eqnarray}
\vert 0_\mathcal{L}\rangle =&&\frac{1}{4}[\vert 00000\rangle+\vert 10010\rangle+\vert 01001\rangle+\vert 10100\rangle\nonumber\\
&&+\vert 01010\rangle-\vert 11011\rangle-\vert 00110\rangle-\vert 11000\rangle\nonumber\\
&&-\vert 11101\rangle-\vert 00011\rangle-\vert 11110\rangle-\vert 01111\rangle\nonumber\\
&&-\vert 10001\rangle-\vert 01100\rangle-\vert 10111\rangle+\vert 00101\rangle],\nonumber
\end{eqnarray}
and
\begin{eqnarray}
\vert 1_\mathcal{L}\rangle=&&\frac{1}{4}[\vert 11111\rangle+\vert 01101\rangle+\vert 10110\rangle+\vert 01011\rangle\nonumber\\
&&+\vert 10101\rangle-\vert 00100\rangle-\vert 11001\rangle-\vert 00111\rangle\nonumber\\
&&-\vert 00010\rangle-\vert 11100\rangle-\vert 00001\rangle-\vert 10000\rangle\nonumber\\
&&-\vert 01110\rangle-\vert10011\rangle-\vert 01000\rangle+\vert 11010\rangle].\nonumber
\end{eqnarray}

We use $H_{\mathcal{S}}$  for the two-dimensional principle system, where the basis vectors are denoted by $\vert 0\rangle$ and $\vert 1\rangle$, while the ancilla system system lies in a $2^4$-dimensional Hilbert space $H_{\mathcal{A}}$ with the basis $\{\vert a_m\rangle\}_{m=0,...,15}$.
The standard way to get the effective noise channel is depicted in Fig.~\ref{fig1}(a). It contains the following steps.

(i) The encoding procedure can be realized with a unitary transformation $U$,
\begin{equation}
U\vert a_0\rangle\otimes \vert 0\rangle\rightarrow\vert 0_\mathcal{L}\rangle,
U\vert a_0\rangle\otimes \vert 1\rangle\rightarrow\vert 1_\mathcal{L}\rangle.
\end{equation}

(ii) The noise evolution is denoted by $\Lambda$.  The five-qubit code above is designed to correct the set of single-qubit errors, $\{E_m\}_{m=0,1,...,15}$. Usually, $E_0$ is fixed to be identity operator $\hat{\texttt{I}}$, and each $E_m(m\neq 0)$ is one of the Pauli operators $\hat{\sigma}^i_j (i=1,...,5, j=x,y,z)$. With the logical codes, one could introduce a set of normalized states
\begin{equation}
\label{norsta}
\vert m,+\rangle=E_m\vert 0_\mathcal{L}\rangle, \vert m,-\rangle=E_m\vert
1_\mathcal{L}\rangle.
 \end{equation}
Since the set of errors, $\{E_m\}_{m=0,...,15}$, could be perfectly corrected, there should be $\hat{P}_{\mathcal{C}}E^{\dagger}_mE_n\hat{P}_{\mathcal{C}}=\delta_{mn}\hat{P}_{\mathcal{C}}$,
where $\hat{P}_{\mathcal{C}}=\vert 0_\mathcal{L}\rangle\langle 0_\mathcal{L}\vert+ \vert
1_\mathcal{L}\rangle\langle
 1_\mathcal{L}\vert$. Therefore, one may easily verify that $\{\vert m, \pm\rangle\}_{m=0}^{15}$ form an orthogonal basis.

 (iii) With the denotation $\vert a_m , i\rangle=\vert a_m\rangle\otimes \vert i\rangle $,  the recovery operation can be described by a process $\mathcal{R}$ such that $\mathcal{R}(\rho^{\mathcal{SA}})=
\sum_{m=0}^{15}R_{m}\rho^{\mathcal{SA}}R_{m}^{\dagger}$, where the Kraus operators $R_m$ are~\cite{Rahn},
\begin{equation}
 R_m=\vert a_m, 0\rangle \langle m, +\vert+\vert a_m, 1\rangle\langle m, -\vert.
 \end{equation}

(iv) The decoding is realized by $U^{\dagger}$, the Hermite conjugate of $U$, $UU^{\dagger}=\mathrm{I}^{\otimes 5}$, and the effective channel $\bar{\varepsilon}$ is
\begin{equation}
\tilde{\varepsilon}(\rho^{\mathcal{S}})=\mathrm{Tr}_{\mathcal{A}}[\mathcal{U}^{\dagger}\circ \mathcal{R}\circ\Lambda\circ \mathcal{U}(\vert a_0\rangle\langle a_0\vert\otimes \rho^{\mathcal{S}})].
\end{equation}

\begin{figure}
\centering
\includegraphics[scale=0.37]{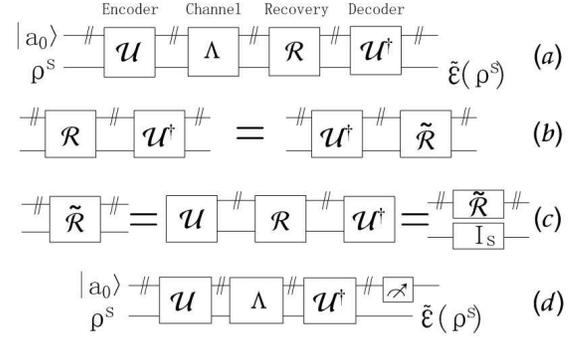}
\caption{(a) The way of getting the effective channel from the standard QEC protocol including encoding, noise evolution, recovery and decoding. (b) The two processes, $\mathcal{R}\circ \mathcal{U}^{\dagger}$ and $\mathcal{U}^{\dagger}\circ \mathcal{\tilde{R}}$, are equivalent.
(c) When  $U^{\dagger}$ for decoding is fixed  according Eq. (6),  the process   $\mathcal{ \tilde{R}}$ can be decomposed as $\mathcal{\tilde{R}}_{\mathcal{SA}}=\mathcal{\tilde{R}}_{\mathcal{A}}\otimes \mathbb{I}_{\mathcal{S}}$. Certainly, it can be moved away.  (d) Our  protocol where the chosen unitary transformation is sufficient to correct the errors of the principle system. The errors of the ancilla system is left to be uncorrected.}
\label{fig1}
\end{figure}

The above is the standard QEC protocol, and the unitary transformation $U$ ($U^{\dagger}$) used for encoding (decoding) is not unique. In this work, however, the unitary transformation is fixed as
\begin{equation}
U\vert a_m\rangle\otimes \vert 0\rangle\rightarrow \vert m, +\rangle, U\vert a_m\rangle\otimes \vert 1\rangle\rightarrow \vert m, -\rangle,\end{equation}
or in the equivalent form
\begin{equation}
\label{udag}
U^{\dagger}\vert m, +\rangle\rightarrow\vert a_m\rangle\otimes \vert 0\rangle,U^{\dagger}\vert m, -\rangle\rightarrow\vert a_m\rangle\otimes \vert 1\rangle,
\end{equation}
Certainly, $\vert 0_\mathcal{L}\rangle =\vert 0,+\rangle, \vert 1_\mathcal{L}\rangle =\vert 0,-\rangle$. [It should be emphasized that the $U^{\dagger}$ introduced  above is nothing else but the $U_2$ used  in the Eq. (87) of the original work in Ref.~\cite {3}.] As it has been argued in Ref.~\cite{3}, the recovery process $\mathcal{R}$ is not a necessary step, since the $U^{\dagger}$ defined in Eq.~(\ref{udag}) is sufficient for correcting the errors of the principle system. One can observe that, the following two processes are equivalent
\begin{equation}
\mathcal{R}\circ \mathcal{U}^{\dagger}\equiv \mathcal{U}^{\dagger}\circ \mathcal{\tilde{R}},
\end{equation}
where the process $\mathcal{\tilde{R}}$ is defined as
\begin{equation}
\mathcal{\tilde{R}}=\mathcal{U}\circ \mathcal{R} \circ \mathcal{U}^{\dagger}.
\end{equation}
Furthermore, it can be expressed with a more explicit way, $\mathcal{\tilde{R}}(\rho^{\mathcal{SA}})=\sum_{m=0}^{15}\tilde{R}_m(\rho^{\mathcal{SA}})\tilde{R}_m^{\dagger}$,
where the Kraus operators $\tilde{R}_m$ take the form
$\tilde{R}_m=UR_mU^{\dagger}$.
By some simple algebra, one may get
$\tilde{R}_m=\vert a_0\rangle\langle  a_m\vert\otimes \mathrm{I}$,
and one can easily verify that: After an arbitrary state $\rho^{\mathcal{SA}}$ of the jointed system is subjected to the process $\mathcal{\tilde{R}}$, the state of the principle system remains unchanged, say,
\begin{equation}
\rho^{\mathcal{S}}=\mathrm{Tr} _{\mathcal{A}}[\rho^{\mathcal{SA}}]=\mathrm{Tr} _{\mathcal{A}}[\mathcal{\tilde{R}}(\rho^{\mathcal{SA}})].
\end{equation}
According to  this  analysis,  the process  $\mathcal{\tilde{R}}$ can be moved away. It has been shown in Fig.~\ref{fig1}(d) that a simplified protocol to obtain the effective channel is defined as,
 \begin{equation}
 \label{vare}
  \tilde{\varepsilon}(\rho^{\mathcal{S}})=\mathrm{Tr} _{\mathcal{A}}[\mathcal{U}^{\dagger}\circ\Lambda\circ \mathcal{U}(\vert a_0\rangle\langle a_0\vert\otimes \rho^{\mathcal{S}})].
  \end{equation}

\section{the standard quantum process tomography}
\label{sec3}
To get the complete information about the effective channel, we shall introduce a convenient tool where a bounded matrix in $H_{2}$ is related to a vector in the enlarged Hilbert space $H_{2}^{\otimes 2}$. Let $A$ be a bounded matrix in the $2$-dimensional Hilbert space $H_2$, with $A_{ij}=\langle i\vert  A\vert j\rangle$
the matrix elements for it, and an isomorphism between $A$ and a $2^2$-dimensional vector $\vert A\rangle\rangle$ is defined as
 \begin{equation}
 \vert A\rangle\rangle =\sqrt{2} A\otimes \mathrm{I}_{2}\vert S_+\rangle=\sum_{i,j=0}^1 A_{ij}\vert ij\rangle,
 \end{equation}
 where $\vert S_+\rangle$ is the maximally entangled state for $H_{2}^{\otimes 2}$, and $\vert S_+ \rangle =\frac{1}{\sqrt{2}}\sum_{k=0}^{1}\vert kk\rangle$ with $\vert ij\rangle=\vert i\rangle\otimes \vert j\rangle$.
This   isomorphism provides a one-to-one mapping between the matrix and its vector form.

For a quantum process $\varepsilon$, the Kraus operators $\{A_m\}$ can be described by a corresponding Choi matrix,
\begin{equation}
\label{choi}
\chi(\varepsilon)=\sum_{m}\vert A_m\rangle\rangle\langle\langle A_m\vert.
\end{equation}
Via the isomorphism above,  this matrix can also be rewritten as
$ \chi(\varepsilon)=2\bar{\rho}$, with $\bar{\rho}=\varepsilon\otimes \mathrm{I}(\vert S_+\rangle\langle S_+\vert)$.
For the normalized state $\bar{\rho}$, Schumacher's entangling  fidelity is defined as $F=\langle S_+\vert\bar{\rho}\vert S_+\rangle$, and it provides a measure for how well the entanglement is preserved by the quantum process $\varepsilon$~\cite{29}. Certainly, one may calculate the entangling fidelity
\begin{equation}
\label{fed}
F(\varepsilon)=\frac{1}{2} \langle S_+\vert\chi(\varepsilon)\vert S_+\rangle.
 \end{equation}

With $\chi$ known, one can derive the Kraus operators of $\varepsilon$. This can be completed through the following simple protocol: The eigenvalues $\lambda^m$ and the corresponding eigenvectors $\vert \Phi^m\rangle$ of $\chi$ can be easily calculated, say,
$\chi=\sum_m\lambda^m\vert \Phi^m\rangle\langle \Phi^m\vert$. Suppose that $\vert\Phi^m\rangle$ can be expanded as $\vert \Phi^m\rangle=\sum_{ij}c^m_{ij}\vert ij\rangle$, with $c^m_{ij}=\langle ij\vert\Phi^m\rangle$ the expanding coefficients. Then, the Kraus operators $A_m$ can be expressed as
\begin{equation}
\label{operator}
A_m=\sqrt{\lambda^m} \sum_{i,j=0}^1 c^m_{ij}\vert i\rangle\langle j\vert.
\end{equation}
With these operators, one may verify that the relation in Eq.~(\ref{choi}) is recovered.

The effective channel can be obtained via the performing the SQPT~\cite{15}. Here, it should be mentioned that the way of performing SQPT is not limited. In the present work, we shall apply the protocol presented in Ref.~\cite{Wu}. For convenience, a brief review of this protocol is organized in following: Introducing the set of operators, say, ${E}_{cd}=\vert c\rangle\langle d\vert (c,d=0,1)$, we take them as the inputs for the principle system, and for a given $E_{cd}$, the corresponding output is
\begin{equation}
\label{varep}
\tilde{\varepsilon}({{E}}_{cd })=\mathrm{Tr}_{\mathcal{A}}[\mathcal{U}^{\dagger}\circ \Lambda\circ \mathcal{U}(\vert a_0\rangle\langle a_0\vert\otimes \vert c\rangle\langle d\vert)].
\end{equation}
Then, one may introduce the coefficients
\begin{equation}
\label{lambda}
\tilde{\lambda}_{ab;cd}=\langle a\vert \tilde{\varepsilon}({{E}}_{cd })\vert b\rangle,
\end{equation}
and $\tilde{\varepsilon}({{E}}_{cd })$ can be expanded as
${\tilde{\varepsilon}}({{E}}_{cd })=\sum_{a,b=0}^{1}\tilde{\lambda}_{ab;cd}\vert a\rangle\langle b\vert$.
The Choi matrix of the effective channel $\tilde{\varepsilon}$ in Eq.~(\ref{vare}), can be expanded as
\begin{equation}
\chi(\tilde{\varepsilon})=\sum_{a,b,c,d=0}^1\tilde{\chi}_{ab;cd} \vert ab\rangle\langle cd \vert,
\end{equation}
with $\tilde{\chi}_{ab;cd}$ its matrix elements,
\begin{equation}
\label{tchi}
\tilde{\chi}_{ab;cd}=\langle ab\vert \chi(\tilde{\varepsilon})\vert cd\rangle.
\end{equation}
It has been shown in Ref.~\cite{Wu} that the $\tilde{\chi}_{ab;cd}$ can be obtained in a simple way,
\begin{equation}
\label{mapping}
\tilde{\chi}_{ab;cd}=\tilde{\lambda}_{ac;bd}.
\end{equation}

\section{exact performance of the concatenated five-qubit code}
\label{sec4}
Now, we shall restrict our attention to the uncorrelated errors. We use $\varepsilon:\{A_m\}$ to denote the quantum process of each two-dimensional  subsystem. Formally, $\Lambda=\varepsilon^{\otimes 5}$. Let us consider the lowest level of error correction for the depolarizing channel $\varepsilon_{\mathrm{DP}}$,
 \begin{equation}
 A_0=\sqrt{F_0}\hat{\texttt{I}}_2, A_i=\sqrt{\frac{1-F_0}{3}}\hat{\sigma}_i
 \end{equation}
 with $i=1,2,3$ and   $F_0$ its channel fidelity. We shall take it as
an explicit example to show how the SQPT is completed.

(a) Let $\vert 0\rangle\langle 0\vert$ the input of the principle system. The corresponding output  is denoted by $\tilde{\varepsilon}(\vert 0\rangle\langle 0\vert)$. With the equation
 \[\tilde{\varepsilon}(\vert 0\rangle\langle 0\vert)=\mathrm{Tr} _{\mathcal{A}}[\mathcal{U}^{\dagger}\circ\Lambda\circ \mathcal{U}(\vert a_0\rangle\langle a_0\vert\otimes \vert 0\rangle\langle 0\vert],\]
one has
\begin{equation}
\tilde{\varepsilon}(\vert 0\rangle\langle 0\vert)=\left(
                                                     \begin{array}{cc}
                                                       a & 0 \\
                                                       0 & 1-a \\
                                                     \end{array}
                                                   \right),
                                                   \nonumber
                                                   \end{equation}
where $a=\frac{1}{81}(1+2F_0)^2(37-108F_0+144F_0^2-64F_0^3)$. Based on the definition in Eq.~(\ref{lambda}),  the four matrix elements, $\tilde{\lambda}_{ab;00}$ ($a,b=0,1$), are
\begin{equation}
\tilde{\lambda}_{00;00}=a, \tilde{\lambda}_{10;00}=\tilde{\lambda}_{01;00}=0, \tilde{\lambda}_{11;00}=1-a.\nonumber
\end{equation}

(b) Similarly with step (a), one may also obtain
\begin{eqnarray}
\tilde{\varepsilon}(\vert 0\rangle\langle 1\vert)&=&\left(
                                                     \begin{array}{cc}
                                                       0 & 2a-1 \\
                                                       0 & 0 \\
                                                     \end{array}
                                                   \right),\nonumber\\
                                                   \tilde{\varepsilon}(\vert 1\rangle\langle 0\vert)&=&\left(
                                                     \begin{array}{cc}
                                                       0 & 0 \\
                                                       2a-1 & 0 \\
                                                     \end{array}
                                                   \right),\nonumber\\
                                                   \tilde{\varepsilon}(\vert 1\rangle\langle 1\vert)&=&\left(
                                                     \begin{array}{cc}
                                                       1-a & 0 \\
                                                       0 & a \\
                                                     \end{array}
                                                   \right),  \nonumber
\end{eqnarray}
and therefore,
\begin{eqnarray}
\tilde{\lambda}_{00;01}= \tilde{\lambda}_{10;01}=\tilde{\lambda}_{11;01}=0, \tilde{\lambda}_{01;01}=2a-1,\nonumber\\
\tilde{\lambda}_{00;10}= \tilde{\lambda}_{01;10}=\tilde{\lambda}_{11;10}=0, \tilde{\lambda}_{10;10}=2a-1,\nonumber\\
\tilde{\lambda}_{00;11}=1-a, \tilde{\lambda}_{10;11}=\tilde{\lambda}_{01;11}=0, \tilde{\lambda}_{11;11}=a.\nonumber
\end{eqnarray}

(c) With all the matrix elements, $\tilde{\lambda}_{ab;cd}$, the so-called matrix $\tilde{\lambda}$ can be orgnized as
 \begin{equation}
 \tilde{\lambda}=\left(
           \begin{array}{cccc}
             a & 0 & 0 & 1-a \\
             0 & 2a-1& 0 & 0 \\
             0 & 0 &  2a-1& 0 \\
             1-a & 0 & 0 & a \\
           \end{array}
         \right).\nonumber
         \end{equation}
 According to the one-to-one relation in  Eq.~(\ref{mapping}), the effective Choi matrix $\tilde{\chi}$ is
  \begin{equation}
\label{varchi}
\tilde{\chi}(\varepsilon_{\mathrm{DEP}})=\left(
               \begin{array}{cccc}
                a& 0 & 0 & 2a-1 \\
                0 & 1-a & 0 & 0 \\
                0 & 0 & 1-a & 0 \\
               2a-1 & 0 & 0 & a \\
                               \end{array}
                                            \right).
  \end{equation}

(d) The entangling fidelity of the effective channel $\tilde{\chi}$ is denoted by $F_1$, and now Eq.~(\ref{fed}) can be rewritten as
\begin{equation}
F_1=\frac{1}{4}(\tilde{\chi}_{00;00}+\tilde{\chi}_{00;11}+\tilde{\chi}_{11;00}+\bar{\chi}_{11;11}).\nonumber
\end{equation}
Based on it, there should be
\begin{equation}
\label{fed1}
F_1(\varepsilon_{\mathrm{DEP}})=\frac{1}{27}(5+20F_0-70F_0^2+40F_0^3+160F_0^4-128F_0^5).
\end{equation}
A parameter $p$ can be used to characterize the depolarizing channel, say, $A_0=\sqrt{1-3p/4}\hat{\texttt{I}}_2, A_i=\sqrt{p}/2\hat{\sigma}_i$, and with the relation $F_0=1-3p/4$, Eq.~(\ref{fed1}) can be rewritten as
\begin{equation}
F_1(\varepsilon_{\mathrm{DEP}})=1-\frac{45}{8}p^2+\frac{75}{8}p^3-\frac{45}{8}p^4+\frac{9}{8}p^5.
\end{equation}
This is the same as the result by Reimpell and Werner~\cite{16}. Furthermore, by requiring that $F_1\ge F_0$, we have the threshold $p<0.18$ (or $F_0>0.86$), the condition under which the five-qubit code works  for the depolarizing channel.

(e) By some simple algebra, the eigenvalues $\lambda^m$ and the corresponding eigenvectors $\vert \Phi^m\rangle$ can be derived,
\begin{eqnarray}
\lambda^0&=&2F_1, \lambda^1=\lambda^2=\lambda^3=\frac{2(1-F_1)}{3},\nonumber\\
\vert\Phi^0\rangle&=&\frac{1}{\sqrt{2}}(\vert 00\rangle+\vert 11\rangle),\nonumber\\
\vert\Phi^1\rangle&=&\frac{1}{\sqrt{2}}(\vert 01\rangle+\vert 10\rangle),\nonumber\\
\vert\Phi^2\rangle&=&\frac{1}{\sqrt{2}}(\vert 01\rangle-\vert 10\rangle),\nonumber\\
\vert\Phi^3\rangle&=&\frac{1}{\sqrt{2}}(\vert 00\rangle-\vert 11\rangle).\nonumber
\end{eqnarray}
With Eq.~(\ref{operator}), one may verify that the effective Choi matrix can be also expressed by a set of Kraus operators $\tilde{A}_m$,
$\tilde{\chi}=\sum_m\vert \tilde{A}_m\rangle\rangle\langle\langle \tilde{A}_m\vert$, where
\begin{equation}
 \tilde{A}_0=\sqrt{F_1}\hat{\texttt{I}}_2, \tilde{A}_i=\sqrt{\frac{1-F_1}{3}}\hat{\sigma}_i
 \end{equation}
 with $i=1,2,3$ and   $F_1$ the entangling fidelity. Obviously, the effective channel
is also a depolarizing channel ~\cite{Rahn,Kersting}.

(f) With the Kraus operators above, a new supper operator $\tilde{\varepsilon}$: $\tilde{\varepsilon}(\rho)=\sum \tilde{A}_m\rho \tilde{A}_m^{\dagger}$ can be defined. Let $\Lambda=\tilde{\varepsilon}^{\otimes 5}$ and follow the steps from (a) to (e), we can obtain the effective
channel by the $5^2$-qubit concatenated code. By repeating the argument above, we can get the effective channel from the $5^L$-qubit
concatenated code.

The protocol developed above, used for the depolarizing channel $ \tilde{\varepsilon}_{\mathrm{DEP}}$, can be easily generalized for other cases. For instance, the amplitude damping channel $\varepsilon_{\mathrm{AD}}$, has been widely discussed in previous works, and the Kraus operators are now
\begin{equation}
\label{damping1}
A_0=\left(
      \begin{array}{cc}
        1 & 0 \\
        0 & \sqrt{1-\gamma} \\
      \end{array}
    \right),A_1=\left(
                  \begin{array}{cc}
                    0 & \sqrt{\gamma} \\
                    0 & 0 \\
                  \end{array}
                \right),\end{equation}
where $\gamma$ is the damping parameter. The entangling fidelity of it is $F_0=\frac{1}{4}(1+\sqrt{1-\gamma})^2$.
 When the five-qubit code is applied for correcting the amplitude damping errors, the entangling fidelity of the effective channel is
\begin{eqnarray}
\label{damping1}
F_1(\gamma)&=&\frac{1}{4}[1+\frac{1}{4}(1-\gamma)^2(4+8\gamma-3\gamma^2+\gamma^3)\nonumber\\
&&+\frac{1}{2}\sqrt{1-\gamma}(4+2\gamma-11\gamma^2+5\gamma^3)].
\end{eqnarray}

Another important case is the bit-flip channel $\varepsilon_{\mathrm{BF}}$ with
\begin{equation}
A_0=\sqrt{F_0}\left(
                \begin{array}{cc}
                  1 & 0 \\
                  0 & 1 \\
                \end{array}
              \right),
 A_1=\sqrt{1-F_0}\left(
                                      \begin{array}{cc}
                                        0 & 1 \\
                                        1 & 0 \\
                                      \end{array}
                                    \right),
                                    \end{equation}
where $F_0$ is the entangling  fidelity.
For a reason which will be clear soon, here we first introduce the denotations,
$\varepsilon_{\mathrm{DP}}\equiv\varepsilon^{0}(\omega_0,F_0)$, $\varepsilon_{\mathrm{AD}}\equiv\varepsilon^{0}(\mathbb{\omega}_1, F_0)$ and $\varepsilon_{\mathrm{BF}}\equiv
\varepsilon^{0}(\omega_2, F_0)$, where $F_0$ is the fidelity of the initial channel without performing QEC. After performing QEC with the
$5^l$-qubit code, the effective channel is denoted by $\varepsilon^{l}(\omega, F_0)$ with  the setting $(\mathbf{\omega}, F_0)$ indicating that the effective channel is originated from the initial $\varepsilon^0(\mathbf{\omega}, F_0)$.
With the Choi matrix $\chi(\varepsilon^{l}(\omega, F_0))$, the entangling fidelity of the effective channel  can be calculated as
\begin{equation}
\label{Fedl}
F_{l}(\omega,F_0)=\frac{1}{2}\langle S_+\vert \chi(\varepsilon^{l}(\omega, F_0))\vert S_+\rangle.
\end{equation}

In the present work, a quantum channel is called \emph{near-perfect} if the entangling fidelity has a value above $1-10^{-5}$. For a given initial channel $\varepsilon^{0}(\omega, F_0)$, we introduce the quantity $L_{(\omega,F_0)}$ and let $5^{L_{(\omega,F_0)}}$ be the minimum size of the concatenated code sufficient for constructing a near-perfect channel, $F_{L_{(\omega,F_0)}}\ge 1-10^{-5}$.
Correspondingly, a $5^{L_{(\omega,F_0)}}$-qubit code is said to be general if $L_{(\omega,F_0)}$ is independent of $\omega$, which is used for denoting the actual type of the error model.

When the concatenated code is applied, the effective  $\chi(\varepsilon^{l}(\omega, F_0))$ ($\omega\neq \omega_0$) usually does not have an analytical form. Noting that $\chi(\varepsilon^{l}(\omega_0, F_0))$   always represents a depolarizing channel, we suppose that $\chi(\varepsilon^{l}(\omega, F_0))$ can be approximated by  $\chi(\varepsilon^{l}(\omega_0, F_0))$, and the error of the approximation is characterized by the distance measure
 \begin{equation}
 \label{dist}
 D_{l}(\omega, F_0)=\frac{1}{4}\vert \chi(\varepsilon^{l}(\mathbf{\omega}, F_0))-\chi(\varepsilon^{l}(\mathbf{\omega}_0, F_0))\vert,
\end{equation}
with $\vert A\vert=\sqrt{A^{\dagger}A}$ ~\cite {15}.  Especially, if $D_{l}(\omega, F_0)\ll 0$,  the entangling fidelities of the two channels,  $\varepsilon^{l}(\mathbf{\omega}, F_0)$ and $\varepsilon^{l}(\mathbf{\omega}_0, F_0)$, almost have the same value.

Under the condition that the entangling fidelity $F_0$ ($F_0\ge 0.86$) is fixed,  besides the steps from (a) to (f) for getting the effective channel, we added another two ones:

(g) With the Choi matrix $\chi(\varepsilon^{1}(\mathbf{\omega}, F_0))$ corresponding to  the effective channel $\varepsilon^{1}(\mathbf{\omega}, F_0)$, the entangling fidelity $F_1( \mathbf{\omega}, F_0)$ is decided by Eq.~(\ref{Fedl}), and the distance $D_{1}(\omega , F_0)$ is calculated   according to Eq.~(\ref{dist}).

(h) With a simple program, the Kraus operators of the effective
channel can be decided, and the calculation for the effective channel in the second level starts from step (a).
 The calculation for a given $\varepsilon(\mathbf{\omega}, F_0)$ can be terminated if
$F_l(\omega, F_0)\geq 1-10^{-5}$.

Based on the iterative protocol developed above, the effective channels in each level of the concatenation can be worked out. For the typical case
where $F_0=0.92$, as shown in Table \Rmnum{1}, we have two observations that: (I) For all the possible channels, a perfect effective channel (with a error below $10^{-5}$) can be constructed by using the same concatenated five-qubit code;
\begin{table}\caption{The fidelity $F_{l}(\omega, F_0)$}
\begin{tabular}{|c|c|c|c|}
  \hline
    $l$ & $F_l(\omega_0)$ & $F_l({\omega_1})$ & $F_{l}(\omega_2)$\\
    \hline
  0 & ~~0.920~~ & ~~0.920~~ & ~~0.920~~ \\
    \hline
  1 & ~~0.946665~~ & ~~0.946762~~ & ~~0.945639~~ \\
    \hline
  2 & ~~0.974784~~ & ~~0.97487~~ & ~~0.973903~~ \\
    \hline
  3 & ~~0.993991~~ & ~~0.99403~~ & ~~0.993576~~ \\
    \hline
  4 & ~~0.999644~~ & ~~0.999648~~ & ~~0.999593~~ \\
    \hline
  5 & ~~0.999999~~ & ~~0.999999~~ & ~~0.999998~~\\
  \hline
\end{tabular}
\end{table}
(II) Meanwhile, the resulted effective channel can be approximated by the depolarizing channel. As shown in Table \Rmnum{2}, the error  of the approximation approaches to zero when $l$, the level of concatenation, is increased.
\begin{table}\caption{The error of the approximation.}
\begin{tabular}{|c|c|c|}
    \hline
  $l$ & $D_l(\omega_1, 0.92)$ & $D_l(\omega_2, 0.92)$ \\
   \hline
 0 & $8.02\times 10^{-2}$ & $5.33\times 10^{-2}$ \\
    \hline
 1 & $3.68\times 10^{-3}$ & $1.79\times 10^{-2}$ \\
   \hline
 2 & $8.77\times 10^{-5}$ & $2.04\times 10^{-3}$ \\
    \hline
 3 & $5.51\times 10^{-6}$ & $2.92\times 10^{-5}$ \\
   \hline
4& $2.15\times 10^{-10}$ & $6.29\times 10^{-9}$ \\
   \hline
 5 & $6.04\times 10^{-14}$ & $1.46\times 10^{-13}$ \\
  \hline
\end{tabular}
\end{table}

The effective channel for other cases, where $F_0$ takes different values, have also been calculated. Our calculation indicates  that  the two observations, (I) and (II) above, are   independent of the choice of $F_0$.

\section{exact performance of the concatenated seven-qubit code}
 \label{section5}
 In this section, we will show that the concatenated seven-qubit code is not general, or in other word, the number of levels for the concatenation, which is necessary for constructing a near-perfect channel, is dependent on the actual type of the noise. For the seven-qubit code~\cite{2}, we can also define a unitary transformation $V$ in a $2^7$-dimensional Hilbert space for encoding, and its inverse $V^{\dagger}$ is applied for decoding.
 Let $\vert 0_{\mathcal{L}}\rangle$ and $\vert 1_{L}\rangle$ be the logical codes, select a set of correctable errors$\{E_m\}_{m=0}^{63}$ including: The identity operator $ E_0=\mathrm{I}^{\otimes 7}$), all the rank-one Pauli operator $\sigma_i^{n}$ ($i=x,y,z$, $n=1,2,...,7)$, and a number of 44 rank-two operators like $\sigma^1_1\otimes \sigma^2_2$, $\sigma^1_3\otimes \sigma^5_2$, ...,\emph{ etc.}, and with the definition
\[ \vert m,+\rangle=E_m\vert 0_{\mathcal{L}}\rangle, \vert m,-\rangle=E_m\vert 1_{\mathcal{L}}\rangle, \] 
we find that the set of normalized vectors $\{\vert m, \pm\rangle\}_{m=0}^{63}$ form the basis of the $2^7$-dimensional Hilbert space.
Use $\{\vert a_m\rangle\}_{m=0}^{63}$ to denote the basis of ancilla system, and with the denotations
$\vert a_m, 0\rangle=\vert a_m \rangle\otimes \vert 0\rangle$ and $\vert a_m, 1\rangle=\vert a_m \rangle\otimes \vert 1\rangle$, the unitary transformation $V$ is
\[V=\sum_{m=0}^{63} \big(\vert m, +\rangle\langle a_m, 0\vert+  \vert m, -\rangle\langle a_m, 1\vert\big).\]
Finally, define $\Lambda=\varepsilon^{\otimes 7}$ and $\vert a_0\rangle=\vert 000000\rangle$, and the effective channel, which is obtained by performing
QEC procedure with the seven-qubit code, can be obtained as
\begin{equation}
  \tilde{\varepsilon}(\rho^{\mathcal{S}})=\mathrm{Tr} _{\mathcal{A}}[\mathcal{V}^{\dagger}\circ\Lambda\circ \mathcal{V}(\vert a_0\rangle\langle a_0\vert\otimes \rho^{\mathcal{S}})].
  \end{equation}

To make sure that our program works in a perfect way, we consider a scenario where the seven-qubit code is applied for the amplitude damping in
Eq.~(\ref{damping1}). The analytical expression for the entangling fidelity of  the
effective channel is
\begin{eqnarray}
\label{damping2}
F_1(\gamma)&=&\frac{1}{4}[1+\sqrt{1-\gamma}(2+\gamma)\nonumber\\
&&+\frac{(1-\gamma)^3}{8}(8+24\gamma-33\gamma^2+21\gamma^3-42\gamma^4)\\
&&+\frac{\sqrt{1-\gamma}}{16}(-150\gamma^2+180\gamma^3-117\gamma^4+39\gamma^5)]\nonumber.
\end{eqnarray}
One can easily check that this result  recovers  the numerical one given in Ref.~\cite{19}, and the entangling fidelities of the effective channel when the five-qubit code and the seven-qubit code are used against the amplitude errors are compared in Fig.~\ref{fig2}.

\begin{figure} \centering
\includegraphics[scale=0.37]{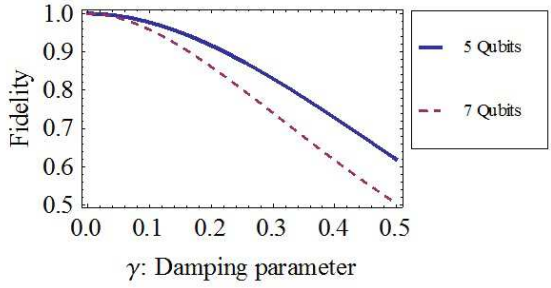}
\caption{(Color online) The entangling fidelities of the effective channel when the five-qubit code and the seven-qubit code are applied against the amplitude damping errors. The exact function in Eq.~(\ref{damping1}) is shown with the solid line while the one  in Eq.~(\ref{damping2}) for the seven-qubit code,  is in the dash line.}
\label{fig2}
\end{figure}

\begin{table}\caption{The fidelity obtained for the concatenated Steane's  code}
\begin{tabular}{|c|c|c|c|}
  \hline
 1& $F_l(\varepsilon_{\mathrm{DEP}})$ & $F_l(\varepsilon_{\mathrm{AD}})$&  $F_l(\varepsilon_{\mathrm{BF}})$ \\  \hline
 0 &0.94 &0.94 & 0.94 \\  \hline
 1 & 0.952211& 0.943496 & 0.943035 \\  \hline
 2 & 0.968897 &0.950234 & 0.947904 \\  \hline
  3& 0.986173 & 0.960975 & 0.955409 \\  \hline
 4 & 0.997048 & 0.975311 & 0.966146 \\  \hline
  5 & 0.99985 & 0.989674 & 0.979469 \\  \hline
 6 & $\geqslant 1-10^{-5}$ & 0.99811 & 0.99196 \\  \hline
  7 & $\geqslant 1-10^{-5}$ & 0.999935 & 0.998693 \\  \hline
 8 & $\geqslant 1-10^{-5}$ & $\geqslant 1-10^{-5}$ & 0.99964 \\  \hline
  9&$\geqslant 1-10^{-5}$ &$\geqslant 1-10^{-5}$&$\geqslant 1-10^{-5}$\\
   \hline
\end{tabular}
\end{table}

As shown in the above section, one can get the exact performance of the QEC protocol based on the concatenated Steane code. For three types of noise models, the depolarizing channel $\varepsilon_{\mathrm{DEP}}$, the amplitude damping channel $\varepsilon_{\mathrm{AD}}$, and
the bit flip channel $\varepsilon_{\mathrm{BF}}$, the corresponding entangling fidelities $F_l$ have been calculated under the condition that the fidelity of the uncorrected channels is fixed to be $F_0=0.94$. Our results, listed in Table~\Rmnum{3}, demonstrate that the necessary numbers of the concatenation are dependend on the actual types of the error.

\section{the general noise model}
\label{sec6}
In this section, we shall show that the observation, which has been observed  for the amplitude damping and bit flip channels, is  general. Let $\{\bar{A}_m\}$ to be a set of Kraus-operators, introduce an arbitrary $2\times2$ unitary transformation,
\begin{equation}
U_2(\theta,\phi)=\left(
                   \begin{array}{cc}
                     \cos\frac{\theta}{2} & \sin\frac{\theta}{2}\exp\{-i\phi\} \\
                    -\sin\frac{\theta}{2}\exp\{i\phi\} & \cos\frac{\theta}{2} \\
                   \end{array}
                 \right),\nonumber
                 \end{equation}
and another set of operators $\{A_m\}$ can be  defined as
\begin{equation}
\label{Abar}
A_m=U_2(\theta,\phi)\bar{A}_m U^{\dagger}_2(\theta,\phi),
\end{equation}
where $\theta$ and $\phi$ are two free parameters.
Three free parameters, $\alpha$, $\beta$, and $\gamma$, are used to define the following four operators,
\begin{eqnarray}
\label{As}
\bar{A}_0=\left(
            \begin{array}{cc}
              \cos\alpha & 0 \\
             0 & \sin\beta\cos\gamma \\
            \end{array}
          \right), \bar{A_1}=\left(
                               \begin{array}{cc}
                                 0 & 0 \\
                                 \sin\alpha\sin\gamma & 0 \\
                               \end{array}
                             \right),\nonumber\\
\bar{A}_2=\left(
            \begin{array}{cc}
              0 & \sin\beta\sin\gamma \\
            0 & 0\\
                       \end{array}  \right),\bar{A}_3=\left(
                                     \begin{array}{cc}
                                       \sin\alpha\cos\gamma & 0 \\
                                     0 & \cos\beta \\
                                     \end{array}
          \right).\nonumber\\
\end{eqnarray}
Now, let us recall some discussions about using the Bloch sphere representation to  describe the single-qubit channel~\cite{15}. With $\vec{r}$ the Bloch vector for an input state $\rho$, and $\vec{\bar{r}}$ for the output state, $\bar{\varepsilon}(\rho)=\sum_m\bar{A}_m\rho(\bar{A}_m)^{\dagger}$, on can obtain a map
\begin{equation}
\vec{r}\rightarrow\vec{\bar{r}}=M\vec{r}+\vec{\delta},
\end{equation}
where $M$ is a $3\times3$ real matrix, and $\vec{\delta}$ is a constant vector. This is an affine map, mapping the Bloch vector into itself. The set of operators in Eq.~(\ref{As}) can be described as
\begin{equation}
\label{map}
\left(
  \begin{array}{c}
    \bar{r}_x \\
    \bar{r}_y \\
    \bar{r}_z \\
  \end{array}
\right)=\left(
          \begin{array}{ccc}
            \eta_{\bot} & 0 & 0 \\
            0 & \eta_{\bot} & 0 \\
            0 & 0 & \eta_z \\
          \end{array}
        \right)\left(
                 \begin{array}{c}
                   r_x \\
                   r_y \\
                   r_z \\
                 \end{array}
               \right)+\left(
                         \begin{array}{c}
                           0 \\
                           0 \\
                           \delta_z \\
                         \end{array}
                       \right),
\end{equation}
with the coefficients
\begin{eqnarray}
\eta_{\bot}&=&\sin(\alpha+\beta)\cos\gamma,\nonumber\\
\eta_z&=&1-(\sin^2\alpha+\sin^2\beta)\sin^2\gamma,\\
\delta_z&=&(\sin^2\beta-\sin^2\alpha)\sin^2\gamma.\nonumber
\end{eqnarray}
 This affine map can be roughly classified into the following two cases: the centered map  with  $\delta_z=0 $ (if $ \alpha=\beta$ ) and the  non-centered one with  $\delta_z\neq 0$.

The centered    map of Eq.~(\ref{map}) is equivalent with the Pauli-channel,
$\sqrt{p_0}\hat{\texttt{I}_2}, \sqrt{p_x}\hat{\sigma}_x,\sqrt{p_y}\hat{\sigma}_y,\sqrt{p_z}\hat{\sigma}_z$, and the parameters $p_i$ are given by
$p_x=p_y=\frac{1}{2}\sin^2\alpha\sin^2\gamma$,
$p_z=\frac{1}{2}(\cos\alpha-\sin\alpha\cos\gamma)^2$,
and $p_0=1-p_x-p_y-p_z$.
 Certainly, it also contains the following two important situations: If $\alpha$ is fixed as
 \begin{equation}
 \cos\alpha=\frac{\cos\gamma+\sin\gamma}{\sqrt{2+\sin 2\gamma}},
 \sin\alpha=\frac{1}{\sqrt{2+\sin 2\gamma}},\nonumber
  \end{equation}
one can obtain a depolarizing channel, while for $\gamma=0$ and an arbitrary $\alpha$, we have the \emph{phase-flip} channel.

With the  $U_2(\theta,\phi)$ introduced above, our noise model should also contain the \emph{bit-flip} channel and \emph{bit-phase flip} channel. For the non-centered case, let $\alpha=0$ and $\beta=\frac{\pi}{2}$, and we can come to the \emph{amplitude damping} channel,
\begin{equation}
\bar{A}_0=\left(
             \begin{array}{cc}
               1& 0 \\
               0 & \cos\gamma \\
             \end{array}
           \right),\bar{A}_1=\left(
                               \begin{array}{cc}
                                 0 & \sin\gamma \\
                               0 &0 \\
                               \end{array}
                             \right).\nonumber
\end{equation}
For the case where the constraint $\alpha+\beta=\frac{\pi}{2}$ holds, we have the so-called \emph{generalized amplitude damping} channel~\cite{15},
\begin{eqnarray}
\bar{A}_0=\cos\alpha\left(
            \begin{array}{cc}
              1 & 0 \\
              0 & \cos\gamma \\
            \end{array}
          \right), \bar{A}_1=\sin\alpha\left(
                                         \begin{array}{cc}
                                           0 & 0 \\
                                           \sin\gamma & 0 \\
                                         \end{array}
                                       \right),\nonumber\\
                                       \bar{A}_2=\cos\alpha\left(
                                                             \begin{array}{cc}
                                                               0 & \sin\gamma \\
                                                               0 & 0 \\
                                                             \end{array}
                                                           \right),\bar{A}_3=\sin\alpha\left(
                                                                                         \begin{array}{cc}
                                                                                           \cos\gamma& 0 \\
                                                                                           0 & 1 \\
                                                                                         \end{array}
                                                                                       \right).\nonumber
 \end{eqnarray}
 From the discussions above, it can be seen that nearly all the noise channel listed in Ref.~\cite{15} are included here. Therefore, our noise model, defined in Eqs.~(\ref{Abar}) and~(\ref{As}), is general.

Now, let us consider a typical case where the entangling fidelity of the uncorrected channel is fixed as $ F_0=0.9$. For the depolarizing channel, using
the result in Eq. (\ref{fed1}),
we can get the entangling fidelities in each level of the concatenation:
\begin{eqnarray}
\label{0.9dep}
F_1(\omega_0)&=&0.920491, F_2(\omega_2)=0.947258, \nonumber \\
F_3(\omega_0)&=&0.975308, F_4(\omega_0)=0.9942310,\nonumber\\
F_5(\omega_0)&=&0.999714, F_6(\omega_0)=0.999999.
\end{eqnarray}
Therefore, to construct a near perfect channel from the depolarizing channel with entangling fidelity $F_0=0.9$, the five-qubit code should be concatenated with itself $L=6$ times. From Eq.~(\ref{varchi}) and the known $F_l(\omega)$, the effective choi matrix
 $\chi(\varepsilon^{1}(\omega_0, F_0))$ can be easily calculated.

 Then,    under the condition that  $F_0$  is fixed, $F_0=0.9$,  we can design a program to generate an arbitrary setting for the five free parameters introduced above. The generated channel is denoted by  $\varepsilon^0(\mathbf{\omega}_1, F_0)$. Let $\Lambda= \varepsilon^0(\mathbf{\omega}_1, F_0)^{\otimes 5}$,
the effective channel in each $l$-th level of concatenation can be decided by following the same method as the amplitude channel in Sec.~\ref{sec4}. As the result of this run of calculation, we can get the exact values $F_l(\omega_1, F_0)$ and $D_l(\omega_1, F_0)$ with $l=1,2,...,6$.
After the calculation of the first one is completed, another channel with a fidelity of $0.9$ will be generated and denoted by $\varepsilon^0(\mathbf{\omega}_2, F_0)$. Similarly, we have the results $F_l(\omega_2, F_0)$ and $D_l(\omega_2, F_0)$ with $l=1,2,...,6$.
Usually, in the $m$-th run of calculation, the generated channel is denoted by $\varepsilon^0(\mathbf{\omega}_m, F_0)$
with $F_0=0.9$. After performing QEC with the $5^6$-qubit concatenated code, we can get a series of exact values
$F_l(\omega_m, F_0)$ and $D_l(\omega_m, F_0)$ with $l=1,2,...,6$.
 For a fixed value of $F_0$, there is about  $M$ ($M\ge 10^5$) examples of noise channel that will be generated.  Based on the numerical data in each level of concatenation, a distance measure $D_l^{\mathrm{max}}(F_0)$ can be defined to denote the maximum value of error for the  approximation defined in Eq.~(\ref{dist}),
\begin{equation}
\label{dist1}
D_l^{\mathrm{max}}(F_0)=\max \{ D_{l}(\omega_m, F_0)\}, 1\leq l\leq 6, 0\leq m\leq M,
\end{equation}
and  the minimum fidelity
\begin{equation}
\label{fedelity}
F_l^{\mathrm{min}}(F_0)=\min \{F_l(\omega_m,F_0)\}_{m=0}^{M}, 1\leq l\leq 6, 0\leq m\leq M.
\end{equation}

For the case $F_0=0.9$, our numerical calculation gives
\begin{eqnarray}
\label{fmin}
F_1^{\mathrm{min}}(F_0)=0.918540, F_2^{\mathrm{min}}(F_0)=0.945006,\nonumber\\
 F_3^{\mathrm{min}}(F_0)=0.973293,
F_4^{\mathrm{min}}(F_0)=0.993281,\nonumber\\ F_5^{\mathrm{min}}(F_0)=0.999555, F_6^{\mathrm{min}}(F_0)=0.999998.
\end{eqnarray}

\begin{figure} \centering
\includegraphics[scale=0.37]{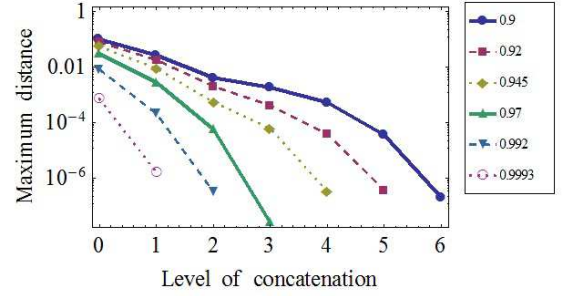}
\caption{(Color online) Numerical results for the maximum distance measure $D_l^{\mathrm{max}}(F_0)$ defined in Eq.~(\ref{dist1}).}
\label{fig3}
\end{figure}

The numerical results for $D^{\mathrm{max}}_l(F_0)$ are given in Fig.~\ref{fig3}, where the initial fidelity take different values, 0.9, 0.92, 0.945, 0.97, 0.97, 0.992, and 0.9993. For each given initial fidelity, the values of the distance measure $D^{\mathrm{max}}_l(F_0)$ are given in the domain $0\leq l\leq L_{(\omega_0, F_0)}$. Our numerical results indicate that: In each level of the concatenation, the effective channel can be approximated by a corresponding depolarizing channel. The error of the approximation, which is characterized by the distance measure $D^{\mathrm{max}}_l(F_0)$, approaches zero as the level of concatenation is increased.

Based on the results in Eq.~(\ref{0.9dep}) and Eq.~(\ref{fmin}), one can observe that: (a) In each level of concatenation, $F_l(\omega_0)-F^{\mathrm{min}}_l\ge 0$. When $l$ is increased, this difference will approach to zero. (b) Since that $F_6^{\mathrm{min}}(F_0)\ge 1-10^{-5}$, the concatenated five-qubit code, which is able to construct a near perfect channel from the depolarizing channel, is also suitable for the arbitrary channel with the same initial fidelity as the depolarizing channel. For the other cases where the initial fidelity takes different values,  0.92, 0.945, 0.97, 0.97, 0.992,
our numerical results, which are depicted in Fig.~\ref{fig4}, show that the properties (a) and (b) can be also observed.

Therefore, as a direct consequence of the observations,  we argue that the concatenated five-qubit is general: Under the condition that the fidelity is fixed,     a  $5^{L_{(\omega_0, F_0)}}$-qubit code, which  is the minimal-sized one for constructing a near-perfect qubit-channel from the  depolarizing channels,  can be used to complete the same task  for the  general noise channels with the same initial fidelity $F_0$. 

\begin{figure} \centering
\includegraphics[scale=0.37]{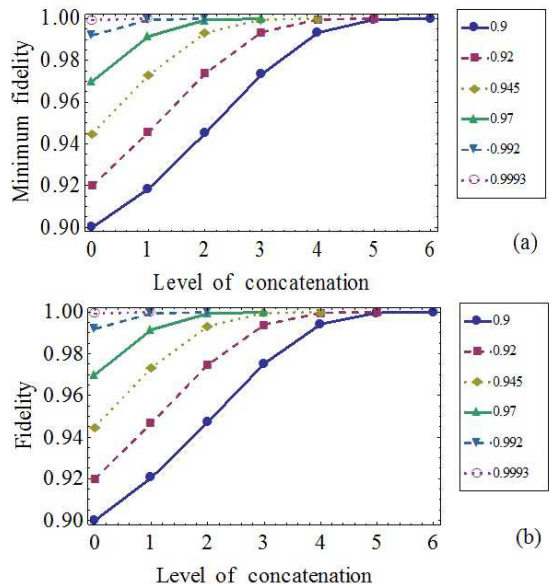}
\caption{(Color online) (a) For the general noise error model, the numerical results of the minimum fidelity $F_l^{\mathrm{min}}(F_0)$ defined in Eq.~(\ref{fedelity}) are depicted; (b) Numerical results for the entangling fidelity when the initial channels are fixed to be the depolarizing ones. The results show the generality of the five-qubit code. Though the numeral results look similar in the two viewgraphs, they are not strictly identical. More specifically, for the case $F_0=0.9$, the entangling fidelities and minimum fidelities in each level of the concatenation given in Eq.~(\ref{0.9dep}) and Eq.~(\ref{fmin}) respectively, are not strictly the same.}
\label{fig4}
\end{figure}

\section{Discussions and remarks}
Our present work is based on the assumption that the noise of quantum channel  comes from the interaction between the physical qubit and the environment. The apparatus, which are used for state  preparation, encoding and decoding, are supposed to be perfect.
The main results of this work are: (I) A simple and explicit scheme is developed to get the effective Choi matrix resulted by the QEC procedure with the concatenated five-qubit code. (II)  Based on this scheme, we  have shown that the QEC procedure, which is optimal for the noise depolarizing channels, also offers an efficient way to construct a near-perfect channel, where the noise of the physical qubit is a general one including the amplitude damping. Within a general noise model, though not a strict proof, our numerical results indicate that the concatenated five-qubit code is general: To construct an  effective channel with a error below $10^{-5}$, the necessary number of the levels for concatenation  is decided by the fidelity of the initial channels and it does not depend on the actual  types of noise models.

Naturally, there is  still an open question: Is the concatenated five-qubit general for an arbitrary noise channel? For the single-qubit case, such a trace-preserving model requires about twelve parameters. Under the constraint that the fidelity is fixed, how to effectively generate an arbitrary setting for all these parameters is still an unsolved problem. From the results of the present work, we guess that the concatenated five-qubit is general. Our  argument is based on the following two facts.

First, a general QEC protocol does exist, and in the work of Horodeckis'~\cite{30}, the so-called twirling procedure has been introduced. Performing twirling on an arbitrary channel $\varepsilon(\omega, F_0)$, one may obtain a depolarizing channel $\varepsilon(\omega_0, F_0)$ with the unchanged channel fidelity, and before performing QEC with the concatenated five-qubit code, all the initial  (unknown) channels may be transferred into the depolarizing channels through twirling.  We may call the so-constructed QEC protocol, which is general for the arbitrary noise models, the \emph{term twirling-assisted QEC scheme}. As a basic property, the effective channel in each level of concatenation should always be a depolarizing channel.

Second, within the general noise model, our numerical results indicate that \emph{the QEC protocol based on the concatenated five-qubit code can be viewed as  an approximate realization of the twirling-assisted QEC scheme}. As shown, the effective channel in each level can be approximated by the depolarizing channel while the error of the approximation approaches to zero as the level of concatenation is increased.

The blockwise decoding protocol, which is suboptimal, is used in our work. From the work of Poulin~\cite{Poulin2}, it is known that the message-passing   decoding algorithm is optimal, and by jointing the twirling stage and the message-passing decoding protocol together, one can have an optimal way to construct a near perfect channel for the arbitrary noise. However, from the results in present work, it is shown that twirling is not a necessary step if the blockwise decoding is applied. Actually, one may guess that the twirling is also not necessary for the QEC with the message-passing decoding algorithm. In other word, the optimal QEC protocol for the depolarizing channel, which has been developed by Poulin, may also be general. We expect this guess could lead to further theoretical or experimental consequences.

\section*{acknowledgements}
The authors are grateful to the referee for very helpful comments. This work was partially supported by the National Natural Science Foundation of China under the Grant No.~11405136, and the Fundamental Research Funds for the Central Universities of China A0920502051411-56.

\end{document}